\input harvmac.tex
%
\figno=0
\def\fig#1#2#3{
\par\begingroup\parindent=0pt\leftskip=1cm\rightskip=1cm\parindent=0pt
\baselineskip=11pt
\global\advance\figno by 1
\midinsert
\epsfxsize=#3
\centerline{\epsfbox{#2}}
\vskip 12pt
{\bf Fig. \the\figno:} #1\par
\endinsert\endgroup\par
}
\def\figlabel#1{\xdef#1{\the\figno}}
\def\encadremath#1{\vbox{\hrule\hbox{\vrule\kern8pt\vbox{\kern8pt
\hbox{$\displaystyle #1$}\kern8pt}
\kern8pt\vrule}\hrule}}

\overfullrule=0pt

%

\def\np#1#2#3{{\it Nucl. Phys.} {\bf B#1} (#2) #3}
\def\pl#1#2#3{{\it Phys. Lett. }{\bf B#1} (#2) #3}
\def\prl#1#2#3{{\it Phys. Rev. Lett.}{\bf #1} (#2) #3}
\def\physrev#1#2#3{{\it Phys. Rev.} {\bf D#1} (#2) #3}

\font\zfont = cmss10 

\def\bigone{\hbox{1\kern -.23em {\rm l}}}
\def\ZZ{\hbox{\zfont Z\kern-.4emZ}}

\def\a{\alpha}

\def\d{\delta}
\def\e{\epsilon}

\def\th{\theta}

\def\r{\rho}

\def\s{\sigma}

\def\c{\chi}

\def\T{\Theta}

\def\O{\Omega}
\def\o{\over}

\def\Gsl{{G }}
\def\Csl{{C }}

\Title{
{\vbox{
\rightline{\hbox{hepth/9911138}}\vskip -0.5mm
\rightline{\hbox{CALT-68-2248}}\vskip -0.5mm  
\rightline{\hbox{CITUSC/99-007}}
}}}
{\vbox{\hbox{\centerline{Fivebrane Gravitational Anomalies }}
\hbox{\centerline{ }}
}}
\smallskip
\centerline{Katrin Becker\footnote{$^1$}
{\tt beckerk@theory.caltech.edu} and Melanie 
Becker\footnote{$^2$}
{\tt mbecker@theory.caltech.edu}} 
\smallskip
\centerline{\it California Institute of Technology 452-48, 
Pasadena, CA 91125}
\smallskip 
\centerline{\it and}
\smallskip
\centerline{\it CIT-USC Center for Theoretical Physics, 
}
\smallskip
\centerline{\it University of Southern California, Los Angeles, CA 90089-2536}

\bigskip
\baselineskip 18pt
\noindent

Freed, Harvey, Minasian and Moore 
have proposed a mechanism to cancel the gravitational anomaly
of the M-theory fivebrane coming from diffeomorphisms acting on the
normal bundle. This procedure is based on a modification of the
conventional M-theory Chern-Simons term. We compactify 
this space-time interaction to the ten-dimensional
type IIA theory. We then analyze the relation to the anomaly cancellation
mechanism for the type IIA fivebrane proposed by Witten.

\Date{November, 1999}

\newsec{Introduction}

In our journey towards M-theory there appears an object that
for a long time has been considered to be rather mysterious, 
the M-theory fivebrane or M5-brane.
The M5-brane was discovered for the first time as a solution
to the classical equations of motion of eleven-dimensional supergravity
\ref\gu{R.~G\"uven, ``Black p-Brane Solutions of $D=11$ Supergravity
Theory'', \pl{276} {1992} {49}.}.
The `fivebrane mystery' originated in part because 
people did not know how to write
down an action for this six-dimensional theory of $(2,0)$ antisymmetric
tensor-multiplets. This problem has been solved in 
\ref\pst{P.~Pasti, D.~Sorokin and M.~Tonin, ``Covariant Action
for a $D=11$ Five-Brane with the Chiral Field'', \pl {398} {1997} {41}, 
hep-th/9701037, 
I.~Bandos, K.~Lechner, A.~Nurmagambetov, P.~Pasti,
D.~Sorokin and M.~Tonin, ``Covariant Action for the Superfive-Brane
of M Theory'', \prl {78} {1997}{4332}, hep-th/9701149.} 
and
\ref\apps{M.~Aganagic, J.~Park, C.~Popescu and J.~Schwarz, 
``Worldvolume Action for the M-Theory Fivebrane'', \np {496}
{1997}{191}, hep-th/9701166.} for the case of one single
fivebrane. However, the action for $N$ coincident 
M5-branes is still not 
known because apparently it is not possible to 
write down the action for a theory of non-abelian tensor-multiplets.
For a recent discussion on the subject see
\ref\bhs{X.~Bekaert, M.~Henneaux and A.~Sevrin, 
``Deformations odd Chiral Two-Forms in Six Dimensions'', 
hep-th/9909094.}. 
Progress towards understanding this system was made 
by Harvey, Minasian and Moore
\ref\hmm{J.~A.~Harvey, R.~Minasian and G.~Moore,
``Nonabelian Tensor Multiplet Anomalies'', JHEP 9809:004, (1998), 
hep-th/9808060.} who computed the gravitational anomalies 
of the non-abelian tensor-multiplet theory through anomaly cancellation
of a system of $N$ coincident M5-branes. The calculation of {\hmm} is based
on the anomaly cancellation mechanism for a single M5-brane proposed
by Freed, Harvey, Minasian and Moore (FHMM) 
\ref\fhmm{D.~Freed, J.~A.~Harvey, R.~Minasian and G.~Moore,
``Gravitational Anomaly Cancellation for M Theory Fivebranes'', 
{\it Adv. Theor. Math. Phys.} {\bf 2} (1998) 601, hep-th/9803205.}. 
Some earlier attempts to understand anomaly cancellation for the 
M5-brane were made in
\ref\al{S.~P.~de Alwis, ``Coupling of Branes and Normalization
of Effective Actions in String/M Theory'', 
\physrev {56} {1997} {7963}, hep-th/9705139.},  
\ref\bcr{L.~Bonora, C.~S.~Chu and M.~Rinaldi, ``Perturbative Anomalies
of the M-5-Brane'', hep-th/9710063; ``Anomalies and Locality 
in Field Theories and M-Theory'', hep-th/9712205.}
and
\ref\he{M.~Henningson, ``Global Anomalies in M Theory'', 
\np {515} {1998} {233}, hep-th/9710126.}.

The perturbative anomaly cancellation for a single fivebrane in type IIA 
theory was
found by Witten
\ref\wi{E.~Witten, ``Five-Brane Effective Action in M-Theory'', 
{\it J. Geom. Phys.} {\bf 22} (1997) 103, 
hep-th/9610234.} who suggested 
an anomaly cancellation mechanism based
on a fivebrane worldvolume counterterm.
Since the type IIA theory is dual to M-theory compactified on a
circle we expect that both anomaly cancellation mechanisms are related
by compactification. It is the purpose of this 
note 
to understand the relation between both mechanisms.
In section 2 we review some basic ideas which one has to keep in
mind for the next sections.
In section 3
we will start with the technically simpler
case of a chiral string in five dimensions. Here the normal bundle
is described by an $SO(3)$ gauge theory. 
This string originates in
compactifications of M-theory on a Calabi-Yau manifold when the M5-brane
wraps a non-trivial four-cycle of the Calabi-Yau. It can potentially have
gauge and gravitational anomalies. We will show that after 
compactification to four dimensions the FHMM Chern-Simons term
gives a string worldvolume counterterm that cancels the four-dimensional 
anomaly. This exactly coincides with the lower-dimensional
version of the anomaly cancellation mechanism proposed in {\wi}.
In section 4 we consider the M5-brane, where 
the normal bundle is described by an $SO(5)$ gauge theory.
The basic idea is the same as in the simpler string example.
After compactification to ten dimensions we will obtain
the fivebrane worldvolume counterterm of {\wi}.
Our conclusions are presented in section 5 and some useful formulas
are collected in an appendix.
In this paper we will study perturbative anomalies. The calculation
of global anomalies in M-theory still remains an open question.

\newsec{The Fivebrane Gravitational Anomaly} 
 
The question that was answered in  
\ref\witt{E.~Witten, ``Five-Branes and M-Theory on an Orbifold'', 
hep-th/9512219.}, {\wi}, {\fhmm} and {\hmm} 
is whether the low energy effective action of M-theory
is well defined even in the presence of extended objects.
The extended objects of M-theory are membranes and fivebranes.
There are no perturbative 
anomalies associated with membrane zero-modes because 
membranes are odd-dimensional and the worldvolume theory is non-chiral.
So the only possible anomalies related to membrane zero-modes are global. 
Their cancellation has been shown in 
\ref\flux{E. Witten, ``Flux Quantization in M-theory'', 
{\it J. Geom. Phys.} {\bf 22} (1997) 1, hep-th/9609122.}.
The situation is very different for the fivebrane because the worldvolume 
is even-dimensional and the worldvolume theory is chiral. Potentially
one could have perturbative anomalies and a calculation is needed to
see if these indeed cancel.
The M5-brane is a six manifold $W_6$ embedded in an eleven
manifold. 
This breaks the Lorentz symmetry to $SO(5,1) \times SO(5)$.
For the low energy effective action of M-theory to be 
well defined, diffeomorphisms which map the fivebrane into itself 
should be a symmetry of the theory.
These are either diffeomorphisms of the fivebrane worldvolume
or $SO(5)$ gauge transformations acting on the normal bundle. 
There are several sources for anomalies. First, there are 
chiral zero-modes that live on the fivebrane worldvolume.
For a single fivebrane these zero-modes form a tensor-multiplet
of six-dimensional $(2,0)$ supersymmetry. 
The chiral fields of this 
multiplet consist of a chiral fermion transforming in the spinor 
representation of $SO(5)$ and a two-form potential with anti-self-dual
field strength which is a singlet under $SO(5)$. The anomaly can be 
calculated with the descent formalism. It is expressed in terms
of an eight-form, $I_8=dI^{(0)}_7$ and the gauge transformation is given
by ${\d}I^{(0)}_7=dI^{(1)}_6$. The anomaly is then given by a
six-form on the fivebrane worldvolume whose explicit form appears in
{\wi}
\eqn\ai{   
2\pi \int_{W_6}I_6^{zm(1)}  .
}

The second source of anomalies comes from Chern-Simons couplings
of the bulk theory 
\eqn\aii{ 
\int_{M_{11}}G_4 \wedge I_7^b.
}
Here $G_4$ is the four-form field strength and 
$I^{b}_7$ is a gravitational Chern-Simons seven-form that can be
expressed in terms of the eleven-dimensional Riemann tensor.
This interaction was discovered in 
\ref\vw{C.~Vafa and E.~Witten, ``A One-Loop Test of String
Duality'', \np {447} {1995} {261}, hep-th/9505053.} by a one-loop 
calculation in type IIA string theory and in
\ref\dlm{M.~J.~. Duff, J.~T.~Liu and R.~Minasian, ``Eleven-
Dimensional Origin of String/String Duality: A One Loop Test'', 
\np {452} {1995} {261}, hep-th/9509084.} 
from anomaly cancellation of the tangent bundle.
In the presence of fivebranes the gauge invariance of this
coupling is broken. 
Under infinitesimal diffeomorphisms $x^I \rightarrow x^I+ \epsilon v^I$
where $\epsilon$ is an infinitesimal parameter and $v$ is a vector
field, $I^{b}_7$ transforms as $I^b_7 \rightarrow I^b_7+d I_6^{b(1)}$,
where $I_6^{b(1)}$ is a six-form. Taking the gauge variation of {\aii}
and integrating by parts we obtain
\eqn\aiii{ 
\delta S=\int_{M_{11}} dG_4 \wedge I_6^{b(1)}.
}
In the presence of fivebranes the four-form is no longer closed
but roughly speaking obeys the Bianchi identity
\eqn\aiv{
dG_4=2\pi \delta_5.
}
Here ${\d}_5$ is a five-form which integrates to one in the
directions transverse to the fivebrane and has a delta function
support on the fivebrane.
The total gravitational anomaly was computed in {\wi} with
some standard formulas appearing in 
\ref\agw{L.~Alvarez-Gaum\'e and E.~Witten, 
``Gravitational Anomalies'', \np {234}{1983}{269}.}
\eqn\av{ 
2\pi \int_{W_6} \left( I_6^{zm}+I_6^b\right)^{(1)}=
2 \pi \int_{W_6} \left( {p_2(N) \over 24 }\right)^{(1)}.
}
Here $p_2(N)$ is the second Pontrjagin class of the normal bundle.
This is in agreement with the result found in {\dlm} who
checked that the anomaly in diffeomorphisms of the tangent bundle
cancel.

A new mechanism was needed to cancel the remaining anomaly.
Until this point there exist two mechanisms in the 
literature that do not seem to have an obvious relationship.
First, Witten {\wi} introduced a counterterm that lives on the fivebrane
worldvolume and that cancels the anomaly in diffeomorphisms of the
normal bundle for the ten-dimensional type IIA theory. 
This mechanism did not seem to have an obvious generalization to 
M5-branes {\wi}.
On the other hand there exists 
the mechanism proposed in {\fhmm} that
is useful for M5-branes. It is based on a 
modification of the eleven-dimensional Chern-Simons term.
Here we will show that both mechanisms are not different after all.
We will see that the compactification of the modified 
eleven-dimensional Chern-Simons term of {\fhmm} to ten dimensions 
gives two contributions. First, a modified ten-dimensional 
Chern-Simons term that is gauge invariant and ensures the compatibility
between the ten-dimensional Bianchi identity and the equations of motion.
Second, the fivebrane worldvolume counterterm proposed
in {\wi} which cancels the ten-dimensional anomaly.
Let us start with the technically simpler case of a chiral string 
in five dimensions which captures the essence of this calculation.

\newsec{The Chiral String or $SO(3)$ Gauge Theory}

Consider a compactification of M-theory on a Calabi-Yau threefold.
When the M5-brane wraps a non-trivial four-cycle of the Calabi-Yau
there appears a chiral string in five dimensions that can have both
gauge and gravitational anomalies.
It was shown in {\fhmm} that in five dimensions these anomalies actually 
cancel. We will show that
after compactification
to four dimensions the FHMM Chern-Simons term {\fhmm} will give  
at least two contributions. 
First, a string
worldvolume counterterm that cancels the gravitational anomaly 
coming from diffeomorphisms of the $SO(2)$ normal bundle.
This worldvolume counterterm is the lower-dimensional version
of the fivebrane worldvolume counterterm found in {\wi}.
Second, a modified four-dimensional Chern-Simons term which is
gauge invariant. This will ensure the compatibility
between the four-dimensional equations of motion and the 
corresponding Bianchi identity.
Let us see how this works in more detail.

\subsec{The String in Five Dimensions}
This case was worked out in {\fhmm} and we will follow closely
their discussion.
Consider a string in five dimensions located at
$y^i=0$ for $i=1,2,3$.
The theory along the string has some properties of 
gravity coupled to an $SO(3)$ gauge theory.
A natural ansatz for the five-dimensional Bianchi identity would be
\eqn\avi{  
d \Gsl_2 =\delta(y_1) \delta(y_2)\d(y_3)  dy^1 \wedge dy^2 \wedge dy^3.  
}
Here $G_2$ is the two-form field strength which couples to the 
string\foot{Here and in the following we absorb a factor $1/{2 \pi}$
in field strengths and potentials.}.
However, the expression {\avi} 
is not well defined when one is dealing 
with interactions that are non-linear in the field strength. 
So for example, order ${\d (0)}^2$-terms may appear
when one is checking the supersymmetry of the Lagrangian.
The appearance of ${\d (0)}^2$-contributions in the Lagrangian was discussed 
some time ago in
\ref\howi{P.~Horava and E.~Witten, ``Eleven-Dimensional Supergravity
on a Manifold with Boundary'', \np {475} {1996} {94}.} 
for compactifications of M-theory on a manifold with boundary.
In fact, there are some striking similarities between both theories.
As in the formula above, the Bianchi identity of {\howi} contains
a delta function whose support is at the boundary 
of space-time. According to {\howi} the appearance of ${\d (0)}^2$-terms
is a symptom of attempting to treat in classical supergravity what really
should be treated in quantum M-theory. In other words, some degrees
of freedom are missing in the classical picture. 
  
In the case at hand the authors of {\fhmm} made a clever proposal in order
to effectively include the missing degrees of freedom. 
FHMM proposed to smooth out the five-dimensional Bianchi identity 
by making the following ansatz
\eqn\avii{
d \Gsl_2=d \rho \wedge {e_2 \over 2}.
}
The right hand side of this identity is the so-called Thom class of the 
normal bundle
\ref\bt{R.~Bott and L.~W.~Tu, ``Differential Form in 
Algebraic Topology'', Springer-Verlag, 1982, New York.}.
$\r=\r(r)$ is a smooth function of the radial direction away from 
the string. It  
takes values $\rho(r)=-1$ for sufficiently small $r$ 
and $\rho(r)=0$ for sufficiently large $r$. 
The delta function of {\avi} 
can be obtained as a limiting function
of the bump form $d\rho$.
The global angular form $e_2$ has integral two over the fiber
and is closed
\eqn\aviii{
de_2=0, 
}
because the Euler class of an odd bundle is zero.
The explicit form of $e_2$ can be found in the appendix of 
{\fhmm}\foot{This corrects a factor 2 in {\fhmm}. This factor has to be 
corrected in the explicit expressions of all angular forms appearing
in {\fhmm} and {\hmm}.}:
\eqn\aix{
e_2={1 \o {4\pi}}\varepsilon_{a b c} 
 \left(D \hat y^a D \hat y^b
 \hat y^c-F^{a b} \hat y^c
\right).
}
Here $\hat y^{a}=y^a/r$ for $a=1,2, 3$ are the coordinates on the 
fiber. Using the connection $\T^{ab}=-\T^{ba}$  
one can define a covariant derivative
\eqn\axi{
D \hat y^a=d\hat y ^a-\Theta^{ab} \hat y^b, 
}
and a two-form 
\eqn\axii{
F^{ab}=d\Theta^{ab}-\Theta^{ac}\wedge \Theta^{cb}. 
}
Formula {\aix} is gauge invariant under the $SO(3)$ gauge
transformations generated by $\a$
\eqn\ax{
{\delta}_{\alpha} \Theta^{ab}=(D \a)^{ab}
\qquad {\rm and} \qquad 
{\delta}_{\alpha} \hat y^{a}=\a^{a b} \hat y^{b}.
}
An equivalent formula for $e_2$ which will turn out to be useful 
for explicit calculations is
\eqn\axiii{
e_2={1 \over 4 \pi}
\varepsilon_{a b c }\left[ 
d \hat y^a d \hat y^b  \hat y^c-d\left( \Theta^{ab} \hat y^c \right) 
\right], 
}
which is identical to the expression for $e_2$ appearing in  
\ref\boca{R.~Bott and A.~S.~Cattaneo, ``Integral invariants 
of 3-Manifolds'', dg-ga/9710001.} that we have collected in the appendix. 

Since the angular form is closed and gauge invariant one can apply
the descent formalism  
\eqn\axiv{
e_2=d e_1^{(0)} \qquad {\rm and} \qquad {\delta}_{\alpha}
e_1^{(0)}=de_0^{(1)}.
}
One can now use the above forms to find the solution to the 
Bianchi identity and the five-dimensional Chern-Simons term.
The solution to the Bianchi identity {\avii} which is non-singular
on the string is given by
\eqn\axvi{
\Gsl_2=d\Csl_1-{1 \over 2} d \rho \wedge e_1^{(0)}.
}
Performing a gauge
transformation and demanding that $\Gsl_2$ should be invariant
one obtains that $\Csl_1$ transforms under gauge transformations of the
normal bundle
\eqn\axvii{
\delta_{\alpha} \Csl_1=-{1 \over 2} {d\rho} \, e_0^{(1)}.
}

In order to have compatibility between the equations of motion and
the Bianchi identity {\avii} one has to correct the 
space-time action. FHMM proposed that one way
of doing this is by introducing a modified Chern-Simons term
\eqn\axv{
S_{CS}={-12 D \pi }
\lim_{\epsilon\rightarrow 0} \int_{M_5 - D_{\epsilon}(W_2) }
\left( \Csl_1-\sigma_1\right)\wedge  d\left( \Csl_1-\sigma_1\right)
\wedge d\left( \Csl_1-\sigma_1\right).
}
Here the integration is over the five-dimensional space-time 
without a tubular region of radius $\e$ around the string, 
$\sigma_1=\rho e^{(0)}_1/2$ and $D$ is a constant 
related to the central charge of the system whose 
precise definition can be found in \ref\msw{J. Maldacena, 
A. Strominger and E. Witten, 
``Black Hole Entropy in M -theory'', hep-th/9711053.} and {\fhmm}.
 
The Chern-Simons interaction is not gauge invariant because
of the gauge transformation
\eqn\axviii{
\delta _{\alpha}(\Csl_1 -\sigma_1)=-{1 \over 2}d(\rho e_0^{(1)}).
}
The variation of the Chern-Simons interaction is then given by a 
boundary term
\eqn\axix{
\delta_{\alpha} S_{CS}= {6 D \pi }
\lim_{\epsilon\rightarrow 0}
\int_{S_{\epsilon} (W_2)}
\rho e_0^{(1)} \left( \Gsl_2 -{1 \o 2} \rho e_2 \right)^2,
}
after applying Stoke's theorem. Here one has to use the fact 
that the boundary of the five-dimensional space without 
the tubular region $D_{\e}(W_2)$ is an $S^2$-sphere bundle over 
the worldvolume of the string $W_2$. This is denoted as $S_{\e}(W_2)$.  
Since $\Gsl_2$ and $\r$ are smooth functions near the brane
one obtains in the limit of small $\epsilon$ 
\eqn\axx{
\delta_{\alpha} S_{CS}=-{3 \o 2} D \pi  
{\int}_{S_{\epsilon}(W_2)}  e_2 e_2 e_0^{(1)} =
-3  D \pi \int_{W_2} p_1^{(1)}(N).
}
The relevant formula used to carry out the remaining integration
is collected in the appendix.
The above anomaly precisely cancels the contributions 
from chiral zero-modes plus anomaly inflow coming from bulk interactions.

\subsec{Compactification to Four Dimensions}

This is the lower-dimensional counterpart of the fivebrane
in ten dimensions that was considered by Witten {\wi}.
The string in four dimensions is described by a  one-form field strength
and a zero-form potential. The FHMM Bianchi identity 
for this case is 
\eqn\bi{
d H_1= { 1\o 2} d(\rho e_1). 
}
$e_1 $ is the  $SO(2)$ invariant global angular form which  
satisfies
\eqn\bii{
de_1 = 2\chi (F), 
}
where $\chi(F)=\epsilon_{ab} F^{ab} /4 \pi$ is the Euler class of the 
$SO(2)$ bundle. 
From {\bi} we see that the Bianchi identity on the brane is given by
\eqn\biii{
d H_1|_{W_2}=-\chi(F),
}
which is equivalent to the Bianchi identity used in {\wi}\foot{
In order to relate the present formulation with the one 
of {\wi} one has to change the sign of $\Theta$. 
This explains the minus sign in front of $\chi(F)$ in 
formula {\biii}. }. 
The explicit form of $e_1$ is
\eqn\biv{
e_1=-{1\over \pi} \epsilon_{ab} (D \hat y )^a \hat y^b=
-{1\over \pi}
\e_{ab}
\left( d \hat y^a \hat y^b-{1\over 2} \T^{ab}\right).
}
In order to make contact with the formalism of {\wi}
we can locally write
\eqn\bixxv{e_1/2=d \Psi_0 +{\Omega}_1,}
where $d \Psi_0$ describes the volume term on the fiber and
\eqn\bv{
\Omega_1={1 \over 4 \pi} \epsilon_{ab} \T^{ab}, 
}
lives on the string worldvolume. This last quantity is the Chern-Simons
one-form in terms of which the Euler character takes the form
$\chi(F)=d \Omega_1$.
Under a gauge transformation we then have 
\eqn\bvi{
\delta_{\alpha} \Omega_1=d \chi_{\alpha}
\qquad {\rm with} \qquad \chi_{\alpha}={1\over 4 \pi}
 \epsilon_{ab} {\alpha}^{ab},  
}
where ${\a}$ is again the generator of the gauge transformation.
Furthermore since $e_1$ is gauge invariant we obtain 
\eqn\bvii{
\delta_{\alpha} \Psi_0=-\chi_{\alpha}. 
}

Using the forms $\Psi_0$ and $\Omega_1$ it is easy to find the
non-singular solution of the Bianchi identity. It is given by
\eqn\bviii{
H_1=dB_0 + \rho \Omega_1- d \rho \Psi_0.
}
Note that a term of the form $\rho e_1$ is not allowed since $e_1$ 
is singular on the brane. To see this note that the integral 
of $e_1$ over the fiber is non-vanishing even if the volume of the fiber
tends to zero. 
Since $H_1$ should be gauge invariant under the $SO(2)$ transformations
of the normal bundle we see that $B_0$ has a transformation
\eqn\bix{
\delta_{\alpha} B_0=- \rho \chi_{\alpha}. 
}

Let us now work out the relation between the $5d$ and $4d$
Chern-Simons terms.
For that purpose we would like to consider an $SO(3)$ bundle $N$
of the form $N=N'\oplus {\cal O}$, where $N'$
is an $SO(2)$ bundle and ${\cal O}$ is a trivial bundle.
We will then assume that the $3$-component of all connections
is equal to zero, $\T^{3 a}=0$ for $a=1,2$. 
The expression for $e_2$ then becomes  
\eqn\bx{
e_2={ 1\over 2 \pi} \omega -\chi (F) \hat y^3
+{ 1\over 4 \pi}\epsilon_{ab} \T^{ab} d \hat y^3 , 
}
where $\omega$ is the volume element of the $SO(3)$ fiber.
Next, notice that the five-dimensional anomaly followed
from the Bott and Cattaneo formula
\eqn\bxiix{
\pi_* e_2^3=2 p_1 (N), 
}
where $\pi_*$ denotes the integration over the fiber. 
Inserting {\bx} into this expression one obtains 
after integration over the fiber
$p_1(N')={\chi(F)}^2$, which is the correct relation between 
the first Pontrjagin class and the Euler class of an even bundle.
The formula that FHMM actually used to compute the 
five-dimensional anomaly is
\eqn\bxii{
\pi_* [e_2 e_2 e_0^{(1)}] =2 p_1^{(1)} (N). 
}
In four dimensions the first Pontrjagin class is
$p_1 (N')={\chi (F)}^2=d({\Omega}_1 \chi(F))$. After
a gauge transformation the right hand side of the previous
equation becomes ${\d}_{\alpha}({\Omega}_1 \chi (F))=
d({\chi}_{\alpha}{\chi})$. This determines $p_1^{(1)} (N') =
{\chi}_{\alpha} {\chi (F)}$.
This is an important piece of information because this together 
with {\bx} and {\bxii} determines the value of
$e_0^{(1)}$
\eqn\bxi{
e_0^{(1)}=-{3\o 2} \chi_{\alpha} \hat y_3.
}
This choice guarantees that the integration over the fiber
correctly reproduces the value of $ p_1^{(1)} (N')$
\eqn\bxii{
\pi_*[ e_2 e_2 e_0^{(1)}] =2 p_1^{(1)} (N')=2 \chi_{\alpha} \chi(F). 
}
Such a choice is always possible 
because $e_1^{(0)}$ is only determined up to a total
derivative. The gauge transformation
of the potential appearing in the Chern-Simons term {\axv} is then 
a total derivative
\eqn\bxiv{
\delta_{\alpha} (C_1 -\sigma_1)=-{ 1\over 2} d( \rho e_0^{(1)})
={ 3\over 4} d( \rho \chi_{\a} \hat y_3). 
}

This can be easily expressed in terms of the gauge transformation
of the four-dimensional $B_0$ field {\bix}, so that we can identify
\eqn\bxv{
C_1-\s_1={ 3 \o 4} d(B_0 \hat y_3)+{\cal I}, 
}
where ${\cal I}$ represents gauge invariant terms. 
The anomalous contribution to the 
Chern-Simons term is then given by
\eqn\bxvii{
S^{\cal A}_{CS}=\lim_{\epsilon\rightarrow 0}
-9 D \pi 
\int_{M_5 - D_{\epsilon ({W_2})} }
d(B_0 \hat y_3) ( \Gsl_2-{1 \o 2} \r e_2 )
( \Gsl_2-{1 \o 2} \r e_2 )}
Using Stoke's theorem we see that $S_{CS}^{\cal A}$ is a boundary term
\eqn\bxviii{
-9 D \pi \lim_{\epsilon\rightarrow 0}
\int_{S_{\epsilon}({W_2}) }
B_0 \hat y_3 ( \Gsl_2-{1 \o 2} \r e_2 )
( \Gsl_2-{1 \o 2} \r e_2 ).
}
Since the fields $B_0$ and $G_2$ 
are smooth near the string 
we obtain after carrying out the limit and the integration over 
the fiber 
\eqn\bxix{
-{9 \o 4} D \pi  \lim_{\epsilon\rightarrow 0}
\int_{S_{\epsilon}({W_2}) }
B_0 e_2 e_2 \hat y_3 = -3 D \pi 
\int _{W_2} B_0 \chi(F) =
3 D \pi \int_{W_2} H_1 \Omega_1. 
}
To carry out the integration over the fiber 
we have used {\bx} and formula (6.6)
of the appendix. 
The result {\bxix} 
is precisely the lower-dimensional analogue of the worldvolume
counterterm 
proposed by Witten for the case of the type IIA fivebrane {\wi}. 

In addition there are four-dimensional space-time interactions 
which are invariant under gauge transformations of the normal
bundle.
We will not determine invariant terms by a direct calculation 
because even in $5d$ the 
complete 
set of invariant terms is not known. This would require a microscopic 
derivation of the interaction
presented by FHMM. This derivation is still missing.
However, in order to have a consistent $4d$ theory we  
expect an additional term in space-time of the form
\eqn\bxx{
\int_{M_{4}} C_1  
(H_1-{1 \o 2} \rho e_1)  G_2. 
}
This can be easily seen because 
once the Bianchi identities are modified $H_1$ is no longer closed.
Replacing $H_1$ by $H_1-\rho e_1 /2$ guarantees that the Bianchi
identities and the equations of motion of this  theory are compatible.

\newsec{The Fivebrane or $SO(5)$ Gauge Theory}

\subsec{The Fivebrane in M-Theory}

The problem of cancelling the anomaly coming from diffeomorphisms
of the normal bundle of an M5-brane was solved in {\fhmm}.
Considering the M5-brane is completely analogous 
to the chiral string in $5d$ that we discussed in the previous section. 
The fivebrane in eleven dimensions breaks the $SO(10,1)$ Lorentz symmetry 
to $SO(5,1) \times SO(5)$. Therefore the normal bundle is described by an 
$SO(5)$ gauge theory.

The M5-brane couples to a 
four-form field strength $G_4$ with a three-form potential $C_3$. 
The FHMM Bianchi identity for this case is
\eqn\ci{
d\Gsl_4=d \rho \wedge { e_4 \over 2}.
}
The explicit form of $e_4$ is given by
\eqn\cii{
\eqalign{
e_4=& {1 \over 32 \pi^2}
\varepsilon_{a_1 \dots a_5}\Bigl[
(D \hat y)^{a_1}
(D \hat y)^{a_2}
(D \hat y)^{a_3}
(D \hat y)^{a_4}\hat y^{a_5} \cr
& -2 F^{a_1 a_2} 
(D \hat y)^{a_3}
(D \hat y)^{a_4}\hat y^{a_5}
+F^{a_1 a_2} F^{a_3 a_4} \hat y^{a_5} \Bigr],}
}
where $a_i, i=1, \dots, 5$ labels the fiber coordinates.
One can again apply the descent formalism and introduce the notations 
\eqn\ciii{
e_4=d e_3^{(0)} \qquad {\rm and} \qquad \delta e_3^{(0)}=d e_2^{(1)}. 
}

The eleven-dimensional FHMM Chern-Simons term is
\eqn\civ{
S_{CS}=-{ 2 \pi \over 6} \lim_{\epsilon\rightarrow 0}
\int_{M_{11} - D_{\epsilon (W_6)} }
(\Csl_3-\s_3)\wedge d(\Csl_3-\s_3)\wedge d(\Csl_3-\s_3), 
}
where $\s_3=\rho e_3^{(0)}/2$. 
This term is not invariant under diffeomorphisms. 
Its variation is obtained by using the anomalous gauge transformation 
of $\Csl_3$ which determines
\eqn\cv{
\d(\Csl_3 -\s_3)=-d( \rho e_2^{(1)} /2).
}
The result for the gauge transformation of the action is then
\eqn\cvi{
\delta S_{CS}=-{  \pi \over 24} 
\int_{S_{\e}(W_6)}  e_4 e_4  e_2^{(1)} =
-{ \pi \o 12} \int_{W_6} p_2^{(1)}(N) .
}
The last identity can be obtained by using the corresponding version 
of the formulas by Bott and Catteneo {\boca}. We have collected 
the relevant 
formulas in the appendix. 

\subsec{ Compactification to Ten Dimensions} 

The anomaly cancellation for the fivebrane in type IIA
theory has been verified in {\wi}.
In ten dimensions the fivebrane is described by a three-form field strength
$H_3$ with a two-form potential $B_2$. The modified Bianchi identity is
\eqn\cvii{
d H_3={1 \over 2} d(\rho e_3). 
}
The angular form $e_3$ is given by
\eqn\cviii{
e_3 =-{1\over 2 \pi^2} \varepsilon_{abcd}\left[
{1 \over 3} (D \hat y)^a 
(D \hat y)^b
(D \hat y)^c \hat y^d-
{ 1\over 2} F^{ab} (D \hat y)^c \hat y^d \right],
}
where $a=1, \dots, 4$ labels the $SO(4)$ fiber coordinates.
Notice that the sign between both terms is different
than in the expression given in {\hmm}. 
This can be written in the form
\eqn\cix{
e_3/2=d\Psi_2 +\O_3, 
}
where $\Psi_2=\Psi_2(\hat y_i, \T)$ is a function of the 
fiber and brane coordinates. It contains besides 
other terms the volume form of $S^4$. Its explicit form 
is not needed in the following. 
$\O_3$ is the Chern-Simons three-form
\eqn\cx{
\O_3=-{ 1\o 32 \pi^2} \e_{abcd} \left(
\T_{ab} d \T_{cd} -{ 2\o 3} \T_{ab} \T_{cx} \T_{xd} \right),
}
which is related to the Euler class by 
\eqn\ccx{
\c(F)=-d \O_3={1\over 32 \pi^2} \varepsilon_{abcd} F^{ab} \wedge 
F^{cd}.
}
Therefore the global angular form $e_3$ is related 
to the Euler class 
\eqn\cxiixi{
de_3=-2\chi(F),
}
as usual. On the brane we the recover 
\eqn\zi{{dH_3}\mid _{W_6}=\chi(F),}
which is the Bianchi identity used in {\wi}.

After an
$SO(4)$ gauge transformation with generator $\a$ 
the Chern-Simons three-form transforms as a total derivative
\eqn\cxi{
\d_{\a} \O_3=d\c_{\a} 
\qquad {\rm with} \qquad 
\c_{\a} =-{ 1\o 32 \pi^2} \e_{abcd} \a^{ab}  d \T ^{cd}.
}
In the notation of {\wi}\foot{There is again a change of sign in $\Theta$.}
 $\O_3=-\O_{\c}(\T)$ and $\c_{\a}=\c(\a,F)$.
The invariance of $e_3$ implies that ${\Psi}_2$ has a gauge
transformation 
\eqn\cxii{
\d_{\a} \Psi_2=-\c_{\a}.
}

In the same way as for the chiral string in $4d$ there is 
only one solution of the Bianchi identity which is non-singular
on the fivebrane
\eqn\cxiii{
H_3=d B_2+  \rho  \O_3- 
d \rho \wedge \Psi_2.
}
Demanding $H_3$ to be gauge invariant we see that 
$B_2$ must have an anomalous variation under 
$SO(4)$ transformations 
\eqn\cxiv{
\d_{\a} B_2=-\rho \c_{\a}.
}
In order to relate the $11d$ and $10d$ theories we will assume 
that the $SO(5)$ bundle $N$ is of the form
$N=N'+{\cal O}$, where $N'$ is an $SO(4)$ bundle and ${\cal O}$
is trivial. The connections involving the five-component
are then vanishing.
Recall that the second Pontrjagin class and the Euler class 
of the $SO(4)$ bundle are related as $p_2(N')=\c(F)^2$, so that we
obtain ${p_2}^{(1)}(N')={\c}_{\alpha}{\c}$.
In order to satisfy the Bott and Cattaneo formula
\eqn\cxv{
\pi_*[e_4 e_4 e_2^{(1)}] =2 {p_2}^{(1)}(N')=2 \chi_{\a} \chi(F), 
}
total derivatives for ${e_3^{(1)}}$ have to be chosen 
in such a way that ${e_2 }^{(1)}$
becomes
\eqn\cxvi{
e_2^{(1)}=45 \c_{\a} \hat y_5.
}
To carry out the integration over the fiber we
have used the formula (6.11) of the appendix.
The $SO(4)$ gauge transformation
of the potential appearing in the eleven-dimensional
Chern-Simons term is then 
\eqn\bxiv{
\delta_{\alpha} (C_3 -\sigma_3)=-{ 1\over 2} d( \rho e_2^{(1)})
=-{ 45\over 2} d( \rho \chi_{\a} \hat y_5). 
}
The anomalous term of the eleven-dimensional Chern-Simons 
interaction can be expressed in terms of the ten-dimensional 
potential $B_2$ as
\eqn\cxviii{
\Csl_3-\s_3={ 45 \over 2} d( B_2 \hat y_5) + {\cal I}, 
}
where ${\cal I}$ is an invariant under $SO(4)$ gauge transformations. 
Because of the appearance of the total derivative in the previous
expression we are able to write the anomalous contribution to the 
Chern-Simons
interaction as a boundary term, exactly as we had done in the lower
dimensional case
\eqn\cxix{
S^{\cal A}_{CS}=-{ 15 \pi \o 2} \lim_{\epsilon \rightarrow 0} 
\int_{S_{\epsilon}(W_6)} \hat y_5 B_2 \wedge (\Gsl_4 -{ 1\o 2} 
\rho e_4)\wedge  (\Gsl_4 -{ 1\o 2} \rho e_4).
}
Since $B_2$ and $\Gsl_4$ are smooth functions near the fivebrane 
the only nonvanishing 
contribution to the above integral is
\eqn\cxx{
-{ 15 \pi \over 8} \lim_{\e\rightarrow 0} 
\int_{S_{\e}(W_6)} B_2 e_4 e_4 \hat y_5.
}
Using formula (6.11) of the appendix we can carry out the
integration over the fiber
\eqn\cxxii{
S_{CS}^{\cal A}=-{\pi \over 12} \int_{W_6} B_2 \c(F)=
{\pi \over 12} \int_{W_6} H_3 \O_3.
}
This is precisely the counterterm found by Witten {\wi}
that cancels the anomaly from $SO(4)$ transformations 
of the normal bundle of a type IIA fivebrane.

Of course, as in the lower-dimensional example, in order 
to have compatibility between the equations of motion
and the Bianchi identity {\cvii} one should have a
modified space-time Chern-Simons term  of the form
\eqn\bxx{
\int_{M_{10} }
 C_3 (H_3-{1 \o 2} \rho e_3) G_4. 
}
However, this does not contribute to the anomaly because it is gauge
invariant. It is very satisfying to see that in ten dimensions this
anomalous contribution to the space-time interaction can indeed be
expressed as a worldvolume counterterm as proposed in {\wi}.

\newsec{Conclusion}
In this paper we have considered a string and a fivebrane
embedded in five and eleven dimensions respectively. 
These theories are not invariant under diffeomorphisms of the normal
bundle. This results in an anomaly that can be expressed 
in terms of the corresponding Pontrjagin class of the normal bundle.
Until now there have been two anomaly cancellation mechanisms 
in the literature whose relation had not been worked out until now.
The mechanism proposed by Freed, Harvey, Minasian and Moore {\fhmm}
is formulated for theories with an odd fiber dimension
while the mechanism proposed by Witten {\wi} is useful for theories
with an even fiber dimension.
In this paper we saw that both mechanism are actually not
different.
We have shown that after compactification 
the FHMM anomaly cancellation mechanism {\fhmm} becomes equivalent
to the one proposed in {\wi}. 
This is very satisfactory and provides further understanding of 
both anomaly cancellation mechanisms.

Even thought a microscopic derivation of the 
FHMM Chern-Simons term is still missing we believe that this interaction
provides a way of 
effectively dealing with the presence of $N$ coincident M5-branes. 
So for example, the proposed interaction 
explains the $N^3$ contributions to the black hole entropy found
in {\msw} from a detailed reduction of the fivebrane tensor-multiplet.
The $N^3$ growth of the entropy of a system of multiple
M5-branes was first discovered in
\ref\kt{I.~R.~Klebanov and A.~Tseytlin, ``Entropy of Near 
Extremal P-Branes'', \np {475} {1996} {164}, hep-th/9604089.}.  

The microscopic description of $N$ coincident fivebranes is 
a theory of non-abelian tensor-multiplets. 
Such a theory does not seem to exist. At least not as 
a local quantum field theory {\bhs}. So the only possibility seems to 
be a non-local theory. 
At this point our most convincing candidate to be a satisfactory 
formulation of M-theory is Matrix theory
\ref\bfss{T.~Banks, W.~Fischler, S.~H.~Shenker and L.~Susskind, 
``M Theory as a Matrix Model: A Conjecture'', 
\physrev {55} {1997} {5112}.}.
However, it is very hard to describe M5-branes in
this approach.  
It is rather possible that the final  
formulation of M-theory may involve a non-local theory. 
Before trying to answer this more difficult
question it would be important to find a global formulation 
of the mechanism proposed by FHMM.
It seems plausible that gerbes 
 \ref\hit{N. Hitchin, ``Lectures on Special Lagrangian
Submanifolds'', math.dg/9907034.} 
might be the correct framework to address this question.
Furthermore, Freed and Witten have computed global anomalies for 
type II theories 
\ref\fw{D.~S.~Freed and E.~Witten, ``Anomalies in String Theory 
with D-Branes'', hep-th/9907189.}. Maybe an extension of this work to 
the eleven-dimensional M-theory is possible.
We hope to report on this in a future.

\vskip 1cm

\noindent {\bf Acknowledgement}

\noindent 

We are grateful to Jeff Harvey, 
Petr Horava, Greg Moore, John Schwarz and Edward Witten 
for useful discussions. 
This work was supported by the U.S. Department of Energy 
under grant DE-FG03-92-ER40701.

\newsec{Appendix}

In this appendix we would like to collect some formulas that are 
useful when computing the integrals involved in the calculation 
of the anomalies presented in this paper. 
We would like to begin with integrals involving $e_2$. 
First $e_2$ can be rewritten in the form
\eqn\api{
e_2={\omega +d( \th^c \hat y_c) \over 2 \pi} 
}
where $\omega$ is the volume element
\eqn\aapi{
\omega=
\hat y_1 d \hat y_2 d\hat y_3 +
\hat y_2 d \hat y_3 d\hat y_1+
\hat y_3 d \hat y_1 d\hat y_2, 
}
which satisfies $\int_{S^2} \omega =4 \pi$ and 
$\th_c=-\e_{abc} \T^{ab}/2$.
This expression implies 
\eqn\apii{
\pi_* e_2 =2
}
Where by $\pi_*$ we denote the integration over the fiber. 
Moreover, 
\eqn\apiii{
\eqalign{
( 2\pi)^2 \pi_* e_2^2  =& \pi_*\left[
2 \omega d(\th^a \hat y_a) +d(\th^a \hat y_a) d(\th^b \hat y_b)\right]=\cr
& 2 d\th^a \pi_*( w \hat y_a) -\th^a \th^b \pi_*( d\hat y_a  d \hat y_b)=0.} 
}
The explicit evaluation of the Bott and Cattaneo formula  
goes along the same lines {\boca} 
\eqn\apiv{
\eqalign{( 2 \pi)^3  \pi_* e_2^3=&
\pi_* \left[ 3 \omega d(\th^a \hat y_a)^2 +d(\th^a \hat y_a)^3 \right]=\cr
& 3 d\th^a d \th^b \pi_*( \omega \hat y_a \hat y_b) 
-3 d\th^a \th^b \th^c \pi_*( \hat y_a d \hat y_b d \hat y_c). } 
}
Introducing spherical coordinates it is the easy to see 
that 
\eqn\apv{
\pi_*( \omega \hat y_a \hat y_b ) ={ 4\over 3} \pi \d_{ab} 
\qquad {\rm and} \qquad \pi_*( \hat y_a d \hat y_b d \hat y_c)
={4 \over 3}\pi \varepsilon_{abc}.
}
Therefore we obtain the equation
\eqn\apvi{
 \pi_* e_2^3 ={ 1\over 2 \pi^2}\left( d \th^a d \th _a -\varepsilon_{abc} d \th^a 
\th^b \th^c\right) =2 p_1.
}
In general $p_r$ denote the Pontrjagin classes. 

The evaluation of these expressions for the $SO(5)$ case is, 
of course, more involved but in principle straightforward. 
Without repeating the same steps as for the $SO(3)$ case we just state 
here the results 
\eqn\apvii{
\pi_* e_4=2, \qquad \pi_* e_4^2 =0 \qquad 
{\rm and} \qquad \pi_* e_4^3=2p_2.
}
If $\omega$ now denotes the volume form of the $S^4$ fiber
\eqn\aapvii{
\omega={ 1 \o 4!} \e_{a_1 a_2 a_3 a_4 a_5} 
d\hat y^{a_1} d \hat y^{a_2} d \hat y^{a_3} 
d \hat y^{a_4 } \hat y^{a_5} , 
}
with $\int_{S^4} \omega =8 \pi^2/3$, then
\eqn\apviii{
\pi_* ( \omega \hat y_a \hat y_b)={ 8 \over 15}\pi^2 \d_{ab}.
}
This equation implies that after imposing the 
condition that the 5-component of the connection vanishes 
we have 
\eqn\apix{
\pi_* (e_4 e_4 \hat y^5) ={ 2\o 45} \c(F), 
}
as a simple evaluation of the involved integrals implies.

\listrefs

\end